\documentclass[12pt]{article}
\usepackage{amsmath}
\usepackage{amssymb}
\usepackage{latexsym}
\usepackage{epsf}
\usepackage{amssymb}

\newtheorem{theorem}{Theorem}
\newtheorem{lemma}{Lemma}
\newtheorem{proposition}{Proposition}

\newtheorem{corollary}{Corollary}
\numberwithin{equation}{section}

\begin{document}

\date{}
\author{M.I.Belishev and M.N.Demchenko}
\title{Time-optimal reconstruction of Riemannian manifold via boundary
electromagnetic measurements} \maketitle
\begin{abstract}
A dynamical Maxwell system is
\begin{align*}
& e_t={\rm curl\,} h, \quad h_t=-{\rm curl\,} e  &&{\rm
in}\,\,\Omega \times (0,T)\\
& e|_{t=0}=0,\,\,\,\,h|_{t=0}=0   &&{\rm in}\,\,\Omega \\
& e_\theta =f &&{\rm in}\,\,\, \partial\Omega \times [0,T]
\end{align*}
where $\Omega$ is a smooth compact oriented $3$-dimensional
Riemannian manifold with boundary, $(\,\cdot\,)_\theta$ is a
tangent component of a vector at the boundary, $e=e^f(x,t)$ and
$h=h^f(x,t)$ are the electric and magnetic components of the
solution. With the system one associates a response operator $R^T:
f \mapsto -\nu \wedge h^f|_{\partial\Omega \times (0,T)}$, where
$\nu$ is an outward normal to $\partial\Omega$.

The time-optimal setup of the inverse problem, which is relevant
to the finiteness of the wave speed propagation, is: given
$R^{2T}$ to recover the part $\Omega^T:=\{x\in \Omega\,|\,{\rm
dist\,}(x,\partial \Omega)<T\}$ of the manifold. As was shown by
Belishev, Isakov, Pestov, Sharafutdinov (2000), for {\it small
enough} $T$ the operator $R^{2T}$ determines $\Omega^T$ uniquely
up to isometry.

Here we prove that uniqueness holds for {\it arbitrary} $T>0$ and
provide a procedure that recovers ${\Omega^T}$ from $R^{2T}$. Our
approach is a version of the boundary control method (Belishev,
1986).
\end{abstract}

\setcounter{section}{-1}
\section{Introduction}
\subsection{Maxwell system}
Let $\Omega$ be a smooth\footnote{Everywhere in the paper,
'smooth' means $C^\infty$-smooth.} compact oriented
three-dimensional Riemannian manifold with the boundary $\Gamma$,
$g$ the metric tensor on $\Omega$. A dynamical (time-domain)
Maxwell system is
\begin{align}
\label{Max1}& e_t={\rm curl\,} h, \quad h_t=-{\rm curl\,} e
&&{\rm
in}\,\,{\rm int\,}\Omega \times (0,T)\\
\label{Max2}& e|_{t=0}=0,\,\,\,\,h|_{t=0}=0   &&{\rm in}\,\,\Omega \\
\label{Max3}& e_\theta =f &&{\rm on}\,\,\, \Gamma \times [0,T]\,,
\end{align}
where ${\rm int\,}\Omega:=\Omega \backslash \Gamma$,
$(\,\cdot\,)_\theta$ is a tangent component of a vector at
$\Gamma$, $f$ is a {\it boundary control}. A solution $\{e,h\}$
describes an electromagnetic wave initiated by the boundary
control, $e=e^f(x,t)$ and $h=h^f(x,t)$ being its electric and
magnetic components. Since the divergence is an integral of
motion, the solution satisfies
\begin{equation}\label{div=0}
{\rm div\,} e^f(\,\cdot\,,t)={\rm div\,} h^f(\,\cdot\,,t)=0 \qquad
\rm in\,\,\, \Omega
\end{equation}
for all $t \geq 0$.

An 'input $\to$ output' correspondence of the system is described
by a {\it response operator} $R^T:f\mapsto -\nu \wedge h^f
|_{{\Gamma \times [0,T]}}$, where $\nu$ is an outward normal at
the boundary, $\wedge$ is the point-wise vector product.

A function ({\it eikonal})
\begin{equation}\label{eikonal}\tau(x):={\rm dist\,}(x, \Gamma)\,, \qquad x \in
\Omega\end{equation} determines the subdomain (near-boundary
layer)
$$\Omega^T:=\{x \in \Omega\,|\,\tau(x)<T\}.$$ By the finiteness of the
domain of influence principle for the Maxwell equations (shortly:
{\it locality principle}), the relation
\begin{equation}\label{supp_eh}
{\rm supp\,}\{e^f, h^f\} \subset  {\{(x,t)\,|\,x\in \Omega,
\,\,\,t\geq \tau(x)\}}
\end{equation}
holds and provides the exact meaning of that the waves propagate
with finite speed.

By the same  principle, the {\it extended problem}
\begin{align}
\label{Maxextended1}& e_t={\rm curl\,} h, \quad h_t=-{\rm curl\,}
e &&{\rm
in}\,\,D^{2T}\\
\label{Maxextended2}& e=0,\quad h=0  &&{\rm
in}\,\,\{(x,t)\in D^{2T}\,|\,t<\tau(x)\} \\
\label{Maxextended3}& e_\theta =f &&{\rm on}\,\,\, \Gamma \times
[0,2T]
\end{align}
in a space-time domain $$D^{2T}:=\{(x,t)\,|\,\,x \in {\rm
int}\,\Omega^T,\,\,\, 0<t<2T-\tau(x)\}$$ turns out to be well
posed, whereas its solution $\{e^f, h^f\}$ is determined by the
part $\Omega^T$ of the manifold. With the problem
(\ref{Maxextended1})--(\ref{Maxextended3}) one associates an {\it
extended response operator} $R^{2T}: f \mapsto  -\nu \wedge
h^f|_{\Gamma \times [0,2T]}$.

\subsection{Main result} So, by the locality principle,
the operator $R^{2T}$ is determined by the part $\Omega^T$ of the
manifold $\Omega$. A reasonable question is: {\bf to what extent
does the operator $R^{2T}$ determine $\Omega^T$?} As was shown in
\cite{DAN}, for small enough $T$'s \footnote{namely, for $T\leq
{\rm dist\,}(c, \Gamma)$, where $c$ is the separation set of
$\Omega$ w.r.t. $\Gamma$} the operator $R^{2T}$ determines
$\Omega^T$ uniquely up to isometry. Here this result is
strengthened as follows.
\begin{theorem}
For any fixed $T>0$, the operator $R^{2T}$ determines the
subdomain $\Omega^T$ up to isometry.
\end{theorem}
The proof is constructive: given $R^{2T}$ we describe a procedure
that provides a manifold $\widetilde \Omega^T$ and endows it with
a metric tensor $\widetilde g$ so that $({\widetilde \Omega}^T,
\,\widetilde g)$ turns out to be isometric to $(\Omega^T,\,g)$.

This result was announced in \cite{BIP07}: as was claimed, it can
be obtained by straightforward generalization of the approach
developed there for the acoustical system to the Maxwell system.
However, the proof, which we propose here, is {\it much simpler}:
it is based on quite elementary geometric facts and the version
\cite{EINT} of the fundamental Holmgren-Joihn-Tataru theorem on
uniqueness of continuation of solutions to the Maxwell system
across a non-characteristic surface. In comparison with the
complicated scheme \cite{BIP07}, which uses such devices as the
Friedrichs extension and Duhamel integral representation, the
reconstruction procedure proposed here looks more prospective for
numerical realization.

\subsection{Comments}
\begin{itemize}
\item Setting the goal to determine an unknown manifold from its
boundary inverse data (here, the response operator $R^{2T}$), we
have to keep in mind the evident nonuiqueness of such a
determination: all {\it isometric} manifolds with the mutual
boundary have the same data. Therefore, the only relevant
understanding of 'to determine' is to construct a manifold, which
possesses the prescribed data \cite{BIP07}. It is what is done in
our paper: we provide the manifold $({\widetilde \Omega}^T,
\,\widetilde g)$, whose response operator $\widetilde R^{2T}$
coincides with the given $R^{2T}$ by construction. \item The
literature devoted to inverse problems of electrodynamics is
hardly observable and we restrict the list of references by the
papers dealing with the {\it time-optimal} setup of the problem
given above. Such an optimality means two things: on the one hand,
just by the locality principle, no $R^{2(T-\varepsilon)}$ with
$\varepsilon>0$ determines $\Omega^T$ and, on the other hand, to
determine $\Omega^T$ we need no $R^{2(T+\varepsilon)}$: it
suffices to know $R^{2T}$. The longer is the time interval of
observations at the boundary, the bigger is the part of the
manifold recovered from the observations. As far as we know, at
the moment the boundary control (BC-) method is the only approach,
which provides such a locality of reconstruction \footnote{In
scalar multidimensional inverse problems, the iterative approach
by V.G.Romanov (see \cite{Rom1}, \cite{Rom2}) is also
time-optimal. Close results for the Maxwell system see in
\cite{RomPukh}.}. \item Dealing with the Maxwell system on a
manifold, we use and refer to the certain facts and results of
\cite{BGlasAA}, which are proved not for this general case but for
the Maxwell system in $\Omega \subset {\mathbb R}^3$ with the
scalar parameters $\varepsilon$ and $\mu$. In all such cases, the
generalization is trivial: to get the proof one can just put
$\varepsilon=\mu \equiv 1$ and use the intrinsic operations ${\rm
curl}, \,\rm div$ etc relevant to the Riemannian structure instead
of the Euclidean one. \item The authors thank Yu.D.Burago and
S.V.Ivanov for very helpful consultations on geometry. We are
grateful to I.V.Kubyshkin for kind help in computer graphics.
\item The work is supported by the RFBR grants No 08-01-00511 and
NSh-4210.2010.1.
\end{itemize}

\section{Geometry}
\subsection{Manifold $\Omega$} We deal with a smooth
compact Riemannian manifold $\Omega$: ${\rm dim~}{\Omega} =3$,
$\Gamma:=\partial \Omega$,  $d$ and $g$ are the distance and
metric tensor in ${\Omega}$. Also, the notation $({\Omega},g)$ is
in use and, if $\Omega$ is considered as a metric space
(regardless its Riemannian structure), we write $({\Omega},d)$.

\noindent{\bf Convention}\,\, In what follows, for a subdomain
$\Sigma \subset \Omega$, by $(\Sigma, \,g)$ we denote this
subdomain endowed with the restriction $g|_\Sigma$ of the metric
tensor. The notation $(\Sigma,\,d)$ means $\Sigma$ endowed with
the {\it interior} distance that is induced by the tensor $g$ and
measured along the curves lying in $\Sigma$.

\noindent Obviously, $(\Sigma, \,g)$ determines $(\Sigma,\,d)$
and, as is well known, the converse is also true: given the
distance $d$ on $\Sigma$ one can recover the smooth structure and
the metric tensor $g$, i.e., determine $(\Sigma,\,g)$.

The {\it eikonal} $\tau(\,\cdot\,): \Omega \to \overline{\mathbb
R}_+$ is defined by (\ref{eikonal}). For a subset $A \subset
{\Omega}$, we denote its metric neighborhood by
$$
{\Omega}^r[A] := \{x \in {\Omega}\, |~d(x, A) < r\}\,, \qquad
r>0\,;
$$
in the case of $A=\Gamma$, we write
$${\Omega}^r:={\Omega}^r[\Gamma]\,=\,\{x \in {\Omega}\, |~\tau(x)
< r\}\,.$$ Later, in dynamics, the value
$$
T_*:=\max_{\Omega} \tau(\cdot)=\inf
\,\{r>0\,|\,{\Omega}^r={\Omega} \}
$$
is interpreted as a time needed for waves moving from $\Gamma$
with the unit speed to fill ${\Omega}$.

The level sets of the eikonal
$$
\Gamma^s:=\{x\in {\Omega}\,|~\tau(x)=s\}\,, \qquad s \geq 0
$$
are the surfaces equidistant to $\Gamma$.
\smallskip

For a subset $A \subset \Omega$, by ${\rm int\,} A$ we denote the
collection of the interior points of the set $A\backslash \Gamma$.

\subsection{Cut locus}
Recall the definition of a separation set (see, e.g, \cite{GKM}).
Let $l_\gamma[0,s]$ be a segment of the length $s$ of a geodesic
$l_\gamma$ emanating from $\gamma \in \Gamma$ orthogonally to
$\Gamma$; let $x(\gamma,s)$ be its second endpoint. The value
$\tau_*(\gamma)$ is said to be a {\it critical length} if
$\tau(x(\gamma,s))=s$ for $0 \leq s \leq \tau_*(\gamma)$ (i.e.,
$l_\gamma[0,s]$ minimizes the distance between $x(\gamma,s)$ and
$\Gamma$) and $\tau(x(\gamma,s))<s$ for $s
>\tau_*(\gamma)$ (i.e., $l_\gamma[0,s]$ does not minimize the distance;
see the illustration on Fig 1, where $s<\tau_*(\gamma)<s^\prime$).
Note that the function $\tau_*(\,\cdot\,)$ is continuous on
$\Gamma$ \cite{GKM}.
\begin{figure}
\centering \epsfysize=7cm \epsfbox{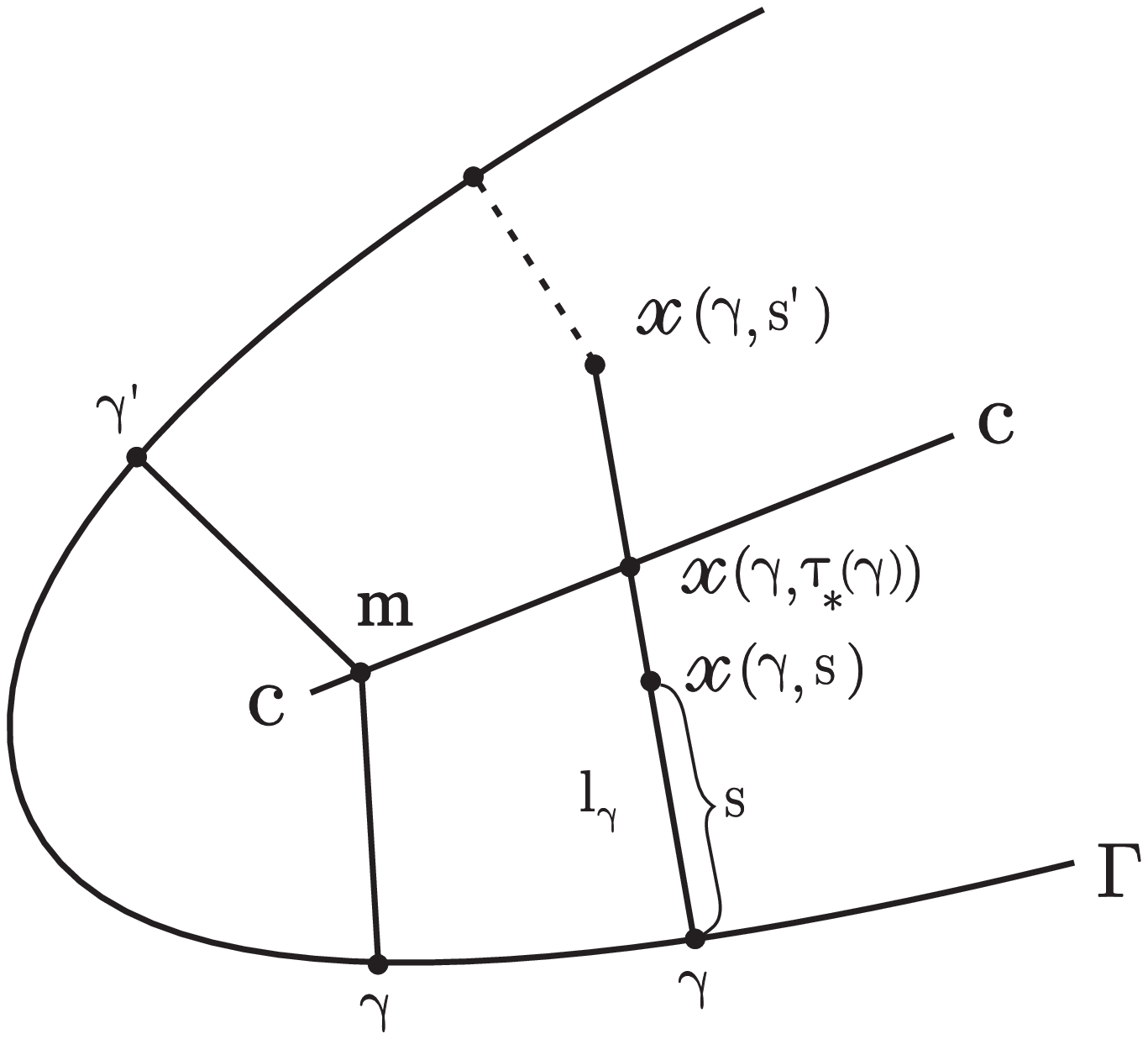}
\caption[Fig.~1]{Cut~locus}
\end{figure}
The point $x(\gamma,\tau_*(\gamma))$ is a {\it separation point}
on $l_\gamma$. A set of the separation points $$c\,:=\, \bigcup
\limits_{\gamma \in \Gamma}x(\gamma,\tau_*(\gamma))$$ is called a
separation set ({\it cut locus}) of $\Omega$ w.r.t. $\Gamma$.

There is one more way to introduce a cut locus. A point $m \in
\Omega$ is said to be {\it multiple} if it is connected with
$\Gamma$ through more than one shortest geodesics (see Fig 1,
where $x(\gamma, s)=x(\gamma',s)=m$ and
$s=\tau(m)=\tau_*(\gamma)=\tau_*(\gamma')$). Denote by $c_0$ the
collection of multiple points and define
$$c\,:=\,\overline c_0.$$ Using (mutatis mutandis) the arguments of
\cite{Kling}, one can check that this definition is equivalent to
the first one .

The cut locus is 'small': the continuity of the function $\tau_*$
easily implies ${\rm vol\,}c\,=\,0$.

Note in addition that $\Gamma^s \backslash c$ is a smooth (may be,
disconnected) surface in ${\Omega}$. If $s<d(c, \Gamma)$ then
$\,\,\Gamma^s$ is smooth and diffeomorphic to $\Gamma$.

\subsection{sgc and pattern}
For any $x \in {\Omega} \backslash c$, there is a unique point
$\gamma(x)\in\Gamma$ nearest to $x$ and a pair
$(\gamma(x),\tau(x))$ is said to be the {\it semigeodesic
coordinates} (sgc) of $x$. If $\gamma^1, \gamma^2$ are the local
coordinates in a neighborhood $\sigma \subset \Gamma$ of
$\gamma(x)$, then the functions
$\gamma^1\left(\gamma(\,\cdot\,)\right),
\gamma^2\left(\gamma(\,\cdot\,)\right), \tau(\,\cdot\,)$
constitute a coordinate system in a 'tube' $\{x \in
\Omega\,|\,\gamma(x) \in
\sigma,\,\,0\leq\tau(x)<\tau_*(\gamma(x))\}$.

A set
$$
\Theta:={\{(\gamma(x),\tau(x))\,|~x \in {\Omega} \backslash c\}}=
\{(\gamma,s)\,|~\gamma \in \Gamma,\,\, 0\leq s < \tau_*(\gamma)\}
\subset \Gamma \times [0,T_*]
$$
is called a {\it pattern} of ${\Omega}$, whereas $$
\theta\,:=\,\bigcup \limits_{\gamma \in
\Gamma}(\gamma,\tau_*(\gamma))\,\subset \Gamma \times [0,T_*]
$$
is its {\it coast} that is the graph of the function $\tau_*$. The
sets
$$\Theta^T:=\Theta \cap \{\Gamma \times [0,T)\}, \qquad \theta^T:=\theta \cap
\{\Gamma \times [0,T)\}$$ are referred to as a pattern of
${\Omega}^T$ and its coast. The patterns are the subgraphs of the
continuous functions $\tau_*(\,\cdot\,)$ and
$\tau_*^T(\,\cdot\,):=\min \left \{ \tau_*(\cdot),T \right \}$:
see Fig.2 \footnote{The illustrations on Fig 2 and 3 are taken
from \cite{BIP07}}.
\begin{figure}
\centering \epsfysize=4cm \epsfbox{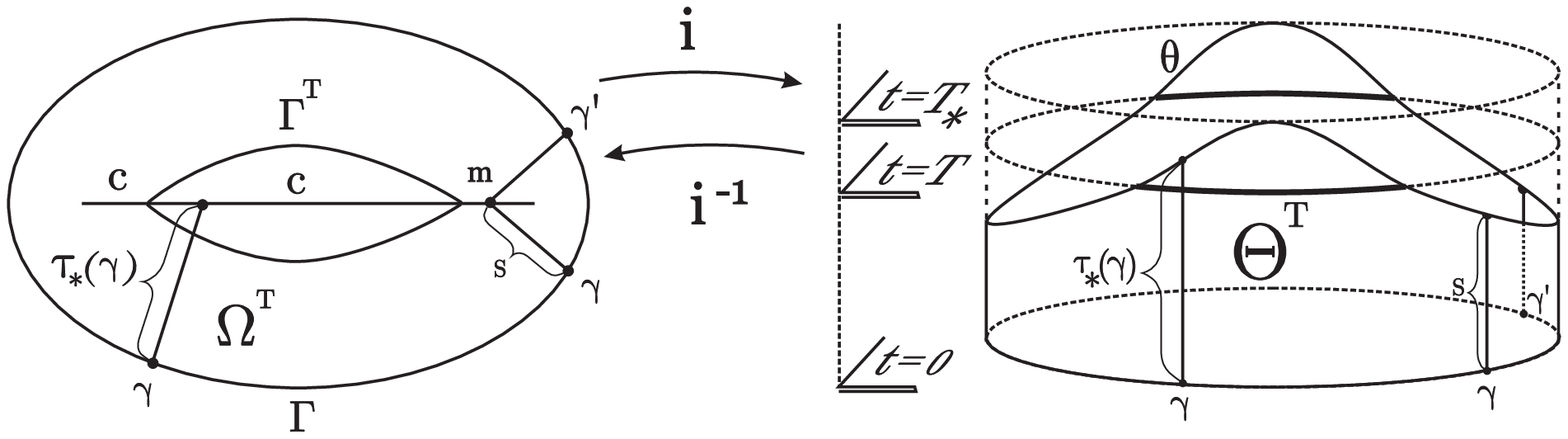}
\caption[Fig.~2]{Pattern}
\end{figure}

A map $i: \Omega \backslash c \to \Theta$,
$$i(x)\,:=\,\left(\gamma(x), \tau(x)\right)$$
is a diffeomorphism\footnote{$\Theta$ is a manifold endowed with
the smooth structure of $\Gamma \times [0, T_*]$.}. Its inverse
$i^{-1}$ transfers $\Theta$ onto $\Omega\backslash c$ by
the rule
$$i^{-1}\left((\gamma,\tau)\right)\,=\,x\left(\gamma,\tau\right)
$$
and can be extended to $\Theta\cup\theta$ by the same rule. In the
sequel, we deal with the extended $i^{-1}$; it maps $\Theta^T \cup
\theta^T$ to $\Omega^T$ continuously and surjectively but not
injectively. As is evident, for two points $(\gamma,s),
(\gamma',s') \in \Theta^T \cup \theta^T$ the equality
\begin{equation}\label{i(-1)(gamma,tau)=m}
i^{-1}\left((\gamma,s)\right)=i^{-1}\left((\gamma',s')\right)\,\,\,(\,=\,m
\in \Omega^T \cap {c})
\end{equation}
is valid iff they lie at the coast $\theta^T$, $s=s'=\tau(m)$
holds, and their mutual image $m$ is a multiple point (see Fig.2).

\subsection{Caps}
Fix $\gamma \in \Gamma$, $s>0$ and a (small) $\varepsilon >0$; let
$\sigma_\varepsilon(\gamma):=\{\gamma^\prime \in
\Gamma|~d(\gamma^\prime,\gamma)<\varepsilon\}$ be a portion of the
boundary. We say a subdomain
\begin{align}
\notag &\omega^{s,\varepsilon}_\gamma\,:={
{\Omega}}^s\left[\sigma_\varepsilon(\gamma)\right]\bigcap
\left\{{{\Omega}}^s \backslash {\Omega}^{s-\varepsilon}\right\}\,=
\\\label{cap}&=\left\{x \in {\Omega}\,|~
d(x,\sigma_\varepsilon(\gamma)) < s, \,\,\,s-\varepsilon \leq
\tau(x) < s\right\}
\end{align}
to be a {\it cap} and note the monotonicity:
$\omega^{s,\varepsilon}_\gamma \subset
\omega^{s,\varepsilon'}_\gamma$ as $\varepsilon<\varepsilon'$.
Introduce a set $$ \lim_{\varepsilon \to
0}\omega^{s,\varepsilon}_\gamma\,:=\,
\bigcap_{0<\varepsilon<s}\overline\omega^{\,s,
\varepsilon}_\gamma~$$ and recall that $x(\gamma,s)$ is defined in
sec.1.2; the following result describes the behavior of the caps
as $\varepsilon \to 0$.
\begin{proposition}\label{Prop2}
The relation
\begin{equation}\label{capslimit}
\lim_{\varepsilon \to 0}\omega^{s,\varepsilon}_\gamma~=~
\begin{cases} \,\,\,\,\,x(\gamma,s)  \qquad &{\rm if}\,\,\,s \leq \tau_*(\gamma) \\
\,\,\,\,\, \emptyset \qquad &{\rm if}\,\,\, s>\tau_*(\gamma)
\end{cases}
\end{equation}
holds.
\end{proposition}
{\bf Proof}\,\, see in \cite{BIP07} (Lemma 1).

So, for a given $\gamma \in \Gamma$ and $s>0$, the cap
$\omega^{s,\varepsilon}_\gamma$ either shrinks to the point
$x(\gamma,s)$ if $(\gamma,s) \in \Theta \cup \theta$, or
terminates (disappears for small enough $\varepsilon$) if
$(\gamma,s) \not \in \Theta\cup \theta$:
\begin{figure}
\centering \epsfysize=7cm \epsfbox{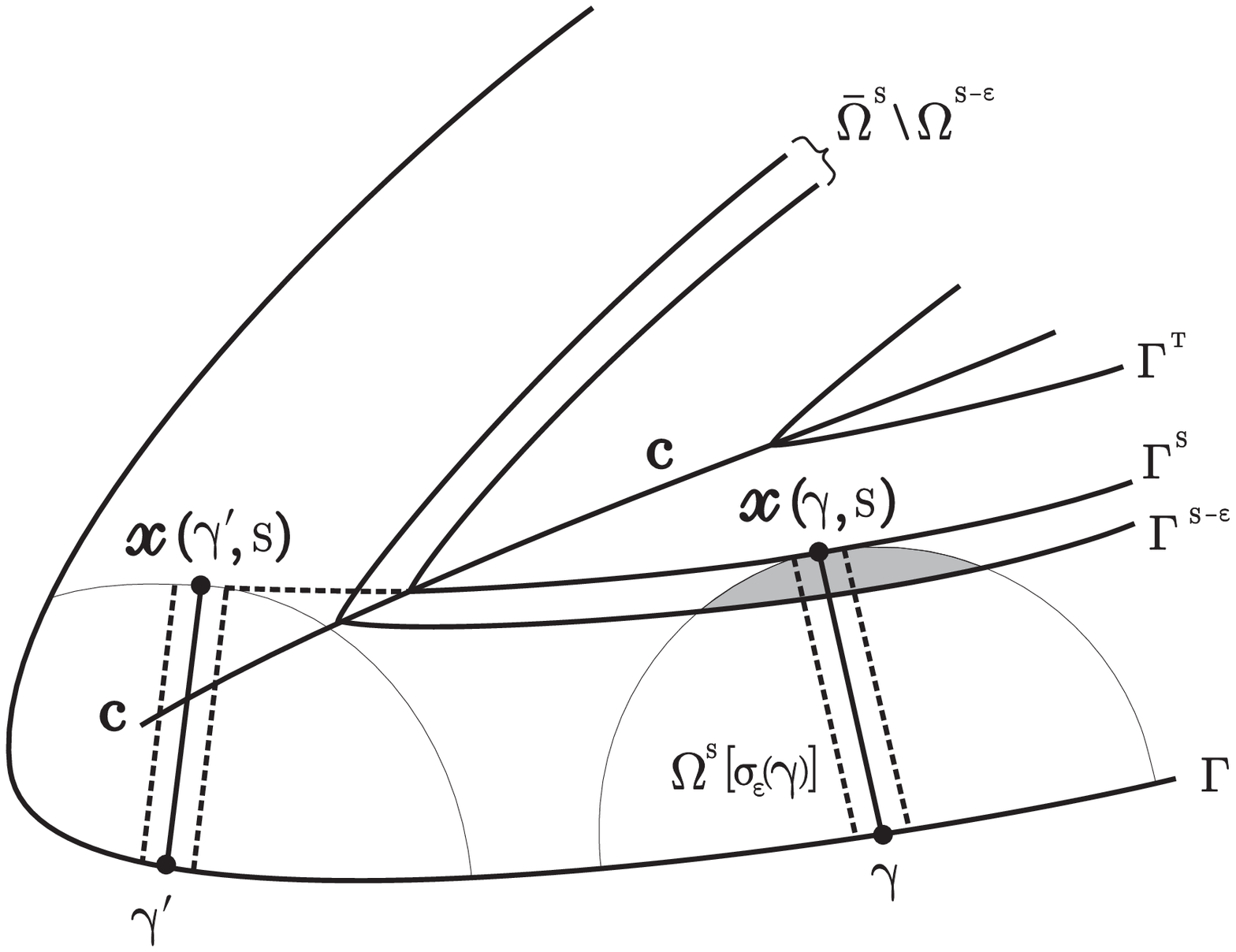}
\caption[Fig.~3]{Cap}
\end{figure}
see Fig 3, where the cap $\omega^{s,\varepsilon}_\gamma$ is
shadowed and $\tau_*(\gamma)>s>\tau_*(\gamma^\prime)$) holds. Such
a behavior of caps leads to the following evident facts, which
we'll use for solving the inverse problem.
\begin{corollary}\label{corollary1}
Let $T>0$ be fixed. A point $(\gamma,s)\in \Gamma \times [0,T)$
belongs to the set $\Theta^T \cup \theta^T$ iff for any
$\varepsilon>0$ the relation
\begin{equation}\label{recoverpattern}
\omega^{s,\varepsilon}_\gamma \not=\emptyset
\end{equation}
holds; in this case, the inequality $s\leq \tau_*(\gamma)$ is
valid. Otherwise, if the family of caps terminates, one has
$(\gamma,s)\notin \Theta^T \cup \theta^T$ and, hence, $s>
\tau_*(\gamma)$ is valid.
\end{corollary}
\begin{corollary}\label{corollary2}
Let $\gamma \in \Gamma$ and $(\gamma',s) \in \Theta^T$, so that
$x'=x(\gamma',s) \in \Omega^T \backslash c$. For a fixed $r<T$,
the inclusion $x' \in \overline \Omega^r[\gamma]$ in $\Omega^T$
(or, equivalently, the inclusion $(\gamma',s) \in i\left(\overline
\Omega^r[\gamma]\backslash c\right)$ on $\Theta^T$) holds iff the
relation
\begin{equation}\label{recoverspheres}
\omega^{s,\varepsilon}_{\gamma'} \bigcap
\,\Omega^{r+\varepsilon}[\sigma_\varepsilon(\gamma)]\,\not=\,\emptyset
\end{equation}
is valid for any $\varepsilon>0$.
\end{corollary}
\begin{corollary}\label{corollary3}
Let the points $(\gamma,s)$ and $(\gamma',s)$ belong to the coast
$\theta^T$. The equality
$i^{-1}\left((\gamma,s)\right)=i^{-1}\left((\gamma',s)\right)$ is
valid iff for any $\varepsilon>0$ the relation
\begin{equation}\label{gluingpattern}\omega^{s,\varepsilon}_\gamma \bigcap\, \Omega^{s+\varepsilon}\left[
\sigma_\varepsilon({\gamma'})\right]\not=\emptyset
\end{equation}
holds (or, equivalently, $\omega^{s,\varepsilon}_{\gamma'} \cap
\Omega^{s+\varepsilon}\left[\sigma_\varepsilon({\gamma})\right]\not=\emptyset$).
\end{corollary}
Corollary \ref{corollary1} is just a convenient reformulation of
Proposition \ref{Prop2}, whereas Corollaries \ref{corollary2} and
\ref{corollary3} easily follow from (\ref{capslimit}) (see the
illustrations on Fig.4a,b, where the caps are shadowed).
\begin{figure}
\centering \epsfysize=7cm \epsfbox{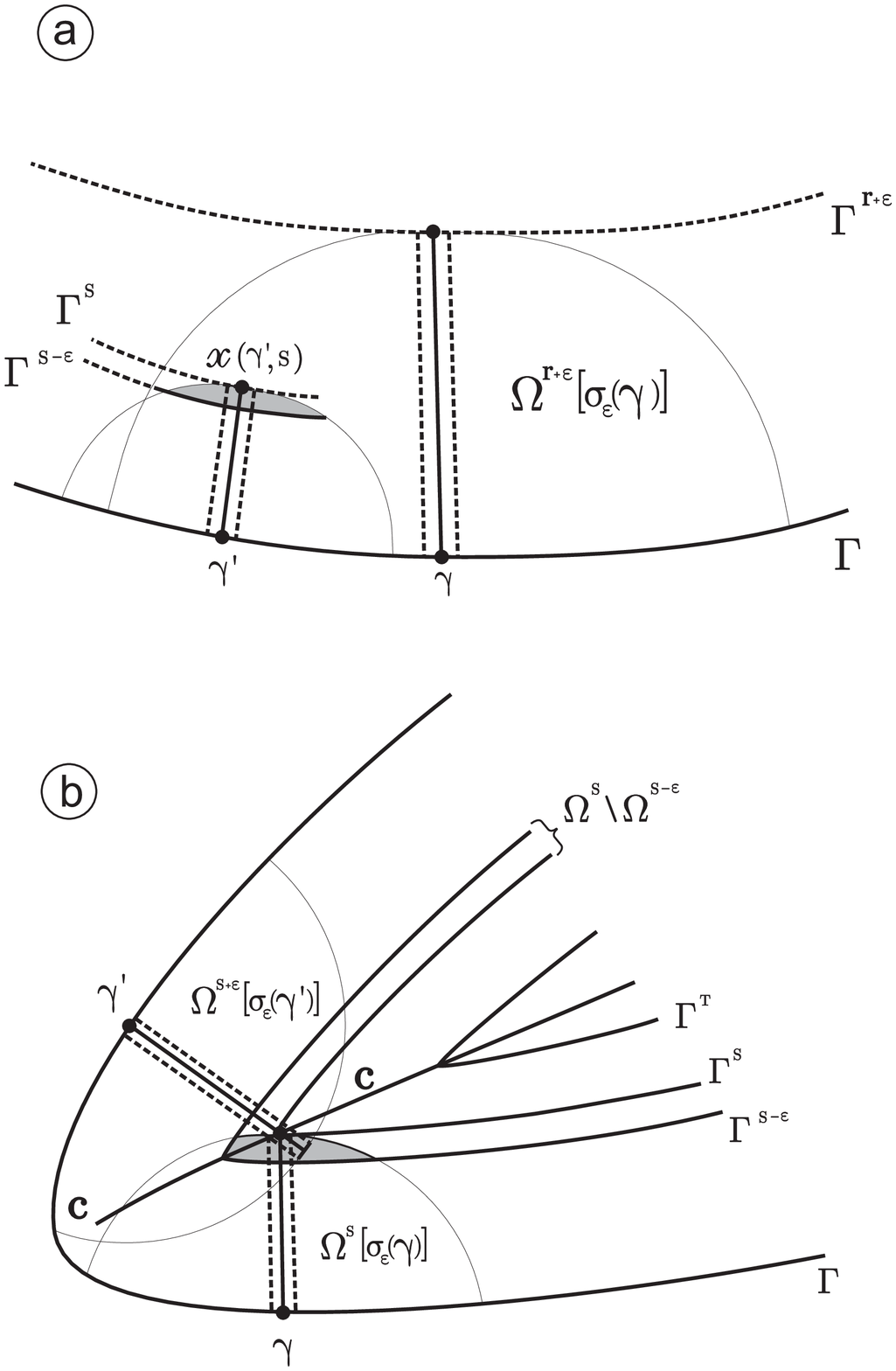}
\caption[Fig.~4]{Relations~(1.5)~and~(1.6)}
\end{figure}

\subsection{Manifold $({\widetilde \Omega}^T, \,\widetilde g)$}
Here we prepare a fragment of the future procedure that solves the
inverse problem. The fragment is the following construction.
\smallskip

Assume that we are given with the pattern $\Theta^T$ endowed with
the tensor $g_{\rm sgc}:=(i^{-1})^*g$ \footnote{in other words,
$g_{\rm sgc}$ is the metric tensor $g$ in the semi-geodesic
coordinates.}. As a metric space, the pair $(\Theta^T, g_{\rm
sgc})$ is isometric to the space $\left(\Omega^T \backslash c,
d\right)$, $d$ being understood as the interior distance
\footnote{Recall the convention of sec. 1.1!}. Our goal is to make
$(\Theta^T, g_{\rm sgc})$ into an isometric copy of
$(\Omega^T,\,g)$.
\begin{itemize} \item{\bf Step 1}\,\,\,Attach the coast $\theta^T$ to
the pattern $\Theta^T$ and extend the metric to $\Theta^T \cup
\theta^T$ by continuity. On the extended space, introduce the
equivalence
\begin{equation}\label{equivalence}
\left\{(\gamma, s)\,\overset{E} =\,(\gamma',
s')\right\}\,\Leftrightarrow\, \left\{i^{-1}(\gamma,
s)\,=\,i^{-1}(\gamma', s')\right\}\,;
\end{equation}
let $\widetilde\Omega^T:=\left[\Theta^T \cup \theta^T\right]
/E\,\,$ be the quotient set, $\pi:\Theta^T \cup \theta^T \to
\widetilde\Omega^T$ the projection. Recalling the aforesaid about
the map $i^{-1}$, we see that the equivalence class of a $(\gamma,
s)\in \Theta^T \cup \theta^T$ consists of more than one element
iff $i^{-1}\left((\gamma, s)\right)$ is a multiple point. Also,
the map $\beta:=i^{-1}\circ\pi^{-1}: \widetilde\Omega^T \to
\Omega^T$ is a well defined bijection. So, as result of sewing the
proper points of the coast, we get a set $\widetilde\Omega^T$ {\it
bijective} to $\Omega^T$. \item {\bf Step 2}\,\,\,Endow
$\widetilde\Omega^T$ with the quotient topology\footnote{$A
\subset \widetilde\Omega^T$ is open iff $\pi^{-1}(A)$ is open in
$\Theta^T \cup \theta^T$.}. As is easy to recognize, the bijection
$\beta$ is a homeomorphism. Thus, we have a topological space
$\widetilde\Omega^T$ {\it homeomorphic} to $(\Omega^T,\,g)$.  It
remains to endow this space with the relevant Riemannian
structure.\item {\bf Step 3}\,\,\, Define a set $\widetilde
c:=\pi(\theta^T) \subset \widetilde\Omega^T$, which is the image
of the cut locus via bijection $\beta^{-1}$. Equip the set
$\widetilde\Omega^T \backslash \widetilde c= \pi(\Theta^T)$ with
the metric tensor $\widetilde g:=(\pi^{-1})^* g_{\rm sgc}$. As is
easy to see, $(\widetilde\Omega^T \backslash \widetilde c,\,
\widetilde g)$ is a manifold isometric to $(\Omega^T \backslash
c,\,g)$.

To extend $\widetilde g$ to $\widetilde c$ one can take a point $a
\in \widetilde c$, its neighborhood $\widetilde \omega \subset
\widetilde\Omega^T$ covered by local coordinates $u^1, u^2, u^3$,
find the matrix $\{\widetilde g_{jk}(u^1, u^2, u^3)\}_{j,k=1}^3$
in $\widetilde \omega \backslash \widetilde c$ and then extend the
matrix elements to $\widetilde \omega \cap \widetilde c$ by
continuity. We denote the extended tensor by the same symbol
$\widetilde g$.
\end{itemize}
As result, we get a Riemannian manifold $({\widetilde \Omega}^T,
\,\widetilde g)$, which is {\it isometric} to $(\Omega^T,\,g)$ by
construction.

\subsection{Tensor $g_{\rm sgc}$ via distant functions}

Return to the starting point of the procedure $(\Theta^T,\, g_{\rm
sgc}) \Rightarrow ({\widetilde \Omega}^T, \,\widetilde g)$ and
explain, where the tensor $g_{\rm sgc}$ will be taken from. Let
$\Omega^r[\gamma]$ be a semi-ball of the radius $r<T$ with the
center $\gamma \in \Gamma$, $i\left(\Omega^r[\gamma]\backslash
c\right) \subset \Theta^T$ its image in the pattern.
\begin{lemma}\label{Lemma1}
The family of the semi-ball images
$${\cal B}^T:=\left\{i\left(\Omega^r[\gamma]\backslash c\right)\,|\,\gamma \in
\Gamma,\,\,0<r<T\right\}$$ determines the tensor $g_{\rm sgc}$ on
$\Theta^T$.
\end{lemma}
{\bf Proof}\,\,\, As is evident, to know the family ${\cal B}^T$
is to know the distant functions
$$r_a(\gamma,\tau)\,:=\,d\left(x(\gamma, \tau), a \right)$$ for all $a \in \Gamma$ and
$(\gamma, \tau) \in \Theta^T$ provided the distance in the r.h.s.
does not exceed $T$.
\smallskip

Fix a point $(\gamma, \tau) \in \Theta^T$. Take a (small)
$\varepsilon
>0$ such that $\eta_{\varepsilon}[\gamma, \tau]:=\sigma_{\varepsilon}(\gamma)
\times (\tau-\varepsilon, \tau+\varepsilon) \subset \Theta^T$ and
$r_a|_{\eta_{\varepsilon}[\gamma, \tau]}<T$ holds for all $a \in
\sigma_{\varepsilon}(\gamma)$. Let $\gamma^1, \gamma^2$ be the
local coordinates on $\sigma_{\varepsilon}(\gamma)$. The metric
tensor $g_{\rm sgc}$ on the pattern $\Theta^T$ (that is the metric
tensor $g$ in sgc) is of the well-known structure
\begin{equation*}
g_{\rm sgc}\,=\,
\begin{pmatrix}
&g_{\rm sgc\,\,11}(\gamma^1, \gamma^2, \tau) && g_{\rm sgc\,\,12}(\gamma^1, \gamma^2, \tau)&&& 0\\
&g_{\rm sgc\,\,21}(\gamma^1, \gamma^2, \tau) && g_{\rm
sgc\,\,22}(\gamma^1, \gamma^2, \tau)&&& 0\\ & 0 && 0 &&& 1
\end{pmatrix}\,;
\end{equation*}
by $\{g_{\rm sgc}^{jk}\}_{j,k=1}^3$ we denote the inverse matrix,
{\it which is of the same structure}.

Choose three {\it base points} $a_i \in
\sigma_{\varepsilon}(\gamma)$ (see Fig.5).
\begin{figure}
\centering \epsfysize=7cm \epsfbox{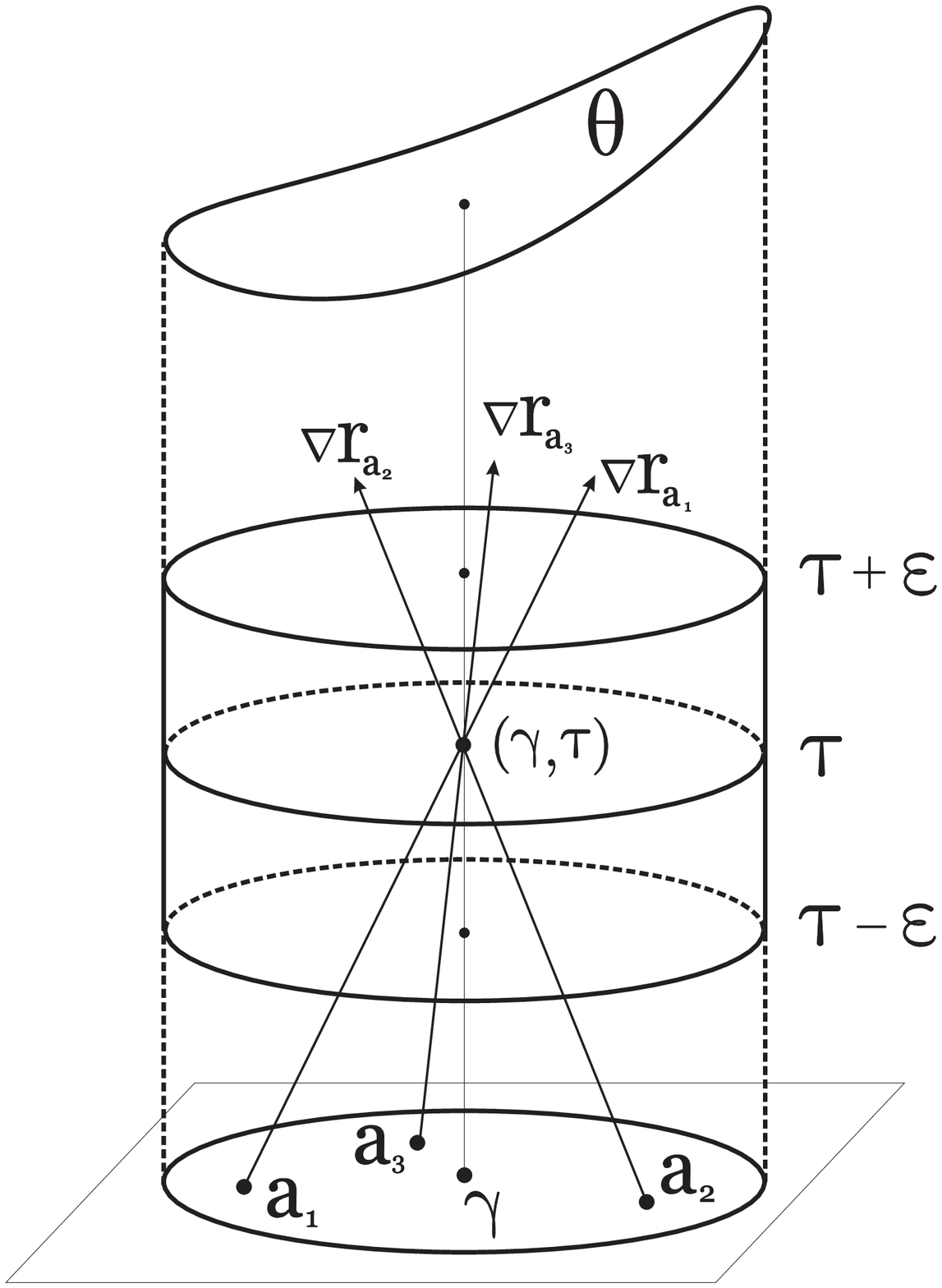}
\caption[Fig.~5]{Recovering $g_{\rm sgc}$}
\end{figure}
The equalities
\begin{equation*}
\left(\frac{\partial r_{a_i}}{\partial \tau}(\gamma,
\tau)\right)^2 + g_{\rm sgc}^{\alpha \beta}(\gamma,
\tau)\,\frac{\partial r_{a_i}}{\partial \gamma^\alpha}(\gamma,
\tau)\,\frac{\partial r_{a_i}}{\partial \gamma^\beta}(\gamma,
\tau)=1\qquad i=1,2,3,\,\,\alpha, \beta =1,2\,,
\end{equation*}
are just the form of writing the well-known fact: the gradient of
{\it any} distant function on $\Omega$ is of the norm $1$.
Consider these equalities as a linear system w.r.t. three unknowns
$g_{\rm sgc}^{11},\,\,g_{\rm sgc}^{12}=g_{\rm sgc}^{21},\,\,g_{\rm
sgc}^{22}$; let $\Delta_{a_1\,a_2\,a_3}(\gamma, \tau)$ be its
determinant. As is easy to show, the freedom in the choice of
$a_1, \,a_2,\,a_3$ is quite enough to provide
$\Delta_{a_1\,a_2\,a_3}(\gamma, \tau)\not=0$. Hence, varying (if
necessary) the position of the base points, one determines from
the system the tensor components at the point $(\gamma, \tau)$.
\,\,$\square$

\subsection{Plan} Here we outline the scheme for solving the inverse
problem.

Assume that the knowledge of the inverse data (operator $R^{2T}$)
enables one to check the relations (\ref{recoverpattern}),
(\ref{recoverspheres}), and (\ref{gluingpattern}). If so, we can
realize the following construction:
\begin{itemize}
\item Select the points $(\gamma,s)\in \Gamma \times [0,T)$, for
which (\ref{recoverpattern}) holds and thus recover the function
$\tau^T_*(\gamma),\,\,\gamma \in \Gamma$, which determines the
pattern $\Theta^T \subset \Gamma \times [0,T)$ and its coast
$\theta^T$ \item Fix $\gamma \in \Gamma$, take $(\gamma',s) \in
\Theta^T,\, r>0$ available for Corollary \ref{corollary2}, and
check by (\ref{recoverspheres}) whether $(\gamma',s)\in
i\left(\overline \Omega^r[\gamma]\backslash c\right)$ holds.
Varying $(\gamma',s)$, recover $i\left(\overline
\Omega^r[\gamma]\backslash c\right)$. Then, varying $\gamma$ and
$r$, determine the family ${\cal B}^T$ of semi-ball images on
$\Theta^T$. By Lemma \ref{Lemma1}, this family determines the
metric tensor $g_{\rm sgc}$ on $\Theta^T$; thus, we get the
manifold $(\Theta^T,\,g_{\rm sgc})$. \item Using
(\ref{gluingpattern}) as a sewing criterion, introduce the
equivalence $E$. Given $(\Theta^T,\,g_{\rm sgc})$ and $E$,
construct the isometric copy $({\widetilde \Omega}^T, \,\widetilde
g)$ of the manifold $(\Omega^T,\,g)$ by the procedure of sec
1.5.\end{itemize}
\smallskip

At this point, one can claim that the inverse problem is solved:
an isometric copy of the original manifold is determined from the
inverse data. It remains to explain, how to check
(\ref{recoverpattern})--(\ref{gluingpattern}) via $R^{2T}$.

\section{Dynamics}
\subsection{Vector analysis} Begin with recalling the definitions
of operations on vector fields in a 3d-manifold (see \cite{Schw}
for detail). As above, $g$ is the metric tensor (2-form on vector
fields). We assume that $\Omega$ is oriented and denote by $\mu$
the volume 3-form.

A {\it scalar product}\, '$\,\cdot\,$' $:\, \{ {\rm fields}\}
\times \{ {\rm fields}\} \, \to \, \{ {\rm functions}\}$ is
defined by $a\cdot b=g(a,b)$.

A {\it vector product} $\wedge :\, \{ {\rm fields}\} \times \{
{\rm fields}\} \, \to \, \{ {\rm fields}\}$ is defined by
$g(a\wedge b,c)=\mu\,(a,b,c)$.

For a field $a$, its {\it conjugate 1-form} $a_\sharp$ is defined
by $a_\sharp(b)=g(a,b)$. Conversely, for an 1-form $\phi$ one
defines its {\it conjugate field} $\phi^\sharp$ by $\phi(b)=g(b,
\phi^\sharp)$.

A {\it gradient} $\nabla :\, \{ {\rm functions}\} \to \{ {\rm
fields}\}$ is $\nabla u=\left({\rm d} u\right)^\sharp$, where $\rm
d$ is the exterior derivative.

A {\it divergence} ${{\rm div\,}}:\, \{ {\rm fields}\} \to \{ {\rm
functions}\}$ acts by ${\rm div\,} a=\star\,{\rm d}\star
a_\sharp$, where $\star$ is the Hodge operator.

A {\it curl} is defined as a map ${\rm curl}:\, \{ {\rm fields}\}
\to \{ {\rm fields}\}$\,,\,~ ${\rm curl\,} a=(\star\,{\rm
d}\,a_\sharp )^\sharp$.
\smallskip

The basic identities are ${\rm div}\,{\rm curl}=0$ and ${\rm
curl}\, \nabla =0$.
\smallskip

Let $\nu$ be the outward unit normal on $\Gamma$, $\,\,
\mu_\Gamma$ the (induced) surface form on $\Gamma$; recall the
Green formula
\begin{equation}\label{Green}
     \int\limits_\Omega u \,{\rm div}\, a\,\, \mu \,=\,
     \int\limits_\Gamma u\, a \cdot \nu\,\, \mu_\Gamma \,-\,
     \int\limits_\Omega a \cdot {\nabla}u\,\,  \mu\,.
\end{equation}

\subsection{System $\alpha^T$}
In what follows, we deal not with the complete Maxwell system
(\ref{Max1})--(\ref{Max3}) but its electric subsystem that is
obtained by eliminating the magnetic component \footnote{A reason
to single out the subsystem is that the components $e^f$ and $h^f$
are not quite independent: for times $t<T_*$ the magnetic
component is determined by electric one \cite{BGlasAA}.}; we write
it in the form
\begin{align}
\label{Electric1} & e_{tt}=-{\rm curl\,}{\rm curl\,} e  &&{\rm
in}\,\,{\rm int\,}\Omega^T \times (0,T)\\
\label{Electric2}& e\,=\,0 &&{\rm
in}\,\{(x,t)\in \Omega^T \times (0,T)\,|\,t<\tau(x)\}\\
\label{Electric3}& e_\theta =f &&{\rm on}\,\,\, \Gamma \times
[0,T]\,,
\end{align}
and refer to as the {\it dynamical system} $\alpha^T$. Such a form
corresponds to the locality principle: the solution $e^f$ is
determined by the part $(\Omega^T,\,g)$ only.

Let us equip $\alpha^T$ with standard control theory attributes:
spaces and operators.
\medskip

\noindent{\bf Outer space}\,\,\,A control $f$ in (\ref{Electric3})
is a time-dependent tangent vector field on $\Gamma$. Let ${\cal
T}$ be a space of square summable tangent fields on $\Gamma$; the
space of controls ${\cal F}^T\,:=\,L_2\left\{[0,T]; {\cal
T}\right\}$ with the product
$$(f,g)_{{\cal F}^T}\,=\,\int\limits_0^T dt \int\limits_{\Gamma }
f(\gamma, t) \cdot g(\gamma, t)\,\,\mu_\Gamma$$ is called {\it
outer}. It contains an extending family of subspaces
$${\cal F}^{\,T,\,\xi}\,:=\,\left\{f \in {\cal F}^T\,|\,\,f|_{0\leq t <T-\xi}=0\right\}\,, \qquad 0\leq \xi \leq T$$
formed by delayed controls; the parameter $\xi$ is regarded as the
action time, whereas $T-\xi$ is the delay.

Also, the class of controls $${\cal F}^T_+:=L_2\left([0,T]; {\cal
T} \cap \vec H^{1 \over 2}(\Gamma)\right)\,,$$ where $\vec H^{1
\over 2}(\Gamma)$ is the Sobolev vector space\footnote{as is
customary on manifolds, we assume $\vec H^{1 \over 2}(\Gamma)$ to
be endowed with one of the equivalent relevant Sobolev norms}, and
the smooth class
$${\cal M}^T:=\left\{f \in C^\infty\left([0,T]; {\cal T} \cap
C^\infty(\Gamma) \right)\,|\,{\rm supp\,} f \subset (0,T]
\right\}$$  are in use. Note that by the definition of the second
class, each $f \in {\cal M}^T$ vanishes near $t=0$. The class
${\cal F}^T_+$ is a normed space w.r.t. the relevant norm, whereas
${\cal M}^T$ is dense in ${\cal F}^T$ and in ${\cal F}^T_+$.

For $f \in {\cal M}^T$, the problem
(\ref{Electric1})--(\ref{Electric3}) has a unique classical
(smooth) solution $e^f$, which is a time-dependent vector field in
$\Omega$. Being initially defined on ${\cal M}^T$, the map $f
\mapsto e^f$ acts continuously from ${\cal F}^T_+$ to the space
$C\left([0,T], {\vec L}_2(\Omega^T)\right)$ (see \cite{Eller}).
Hence, it can be extended to the class ${\cal F}^T_+$. In what
follows we assume that such an extension is done and regard its
images as {\it generalized solutions} of the problem
(\ref{Electric1})--(\ref{Electric3}).
\medskip

\noindent{\bf Inner space}\,\,\,In control theory, the solution
$e^f$ is referred to as a {\it trajectory} of the system
$\alpha^T$, whereas $e^f(\,\cdot\,,t)$ is a {\it state} at the
moment $t$. By (\ref{div=0}), the states are divergence-free
fields, whereas by (\ref{supp_eh}) one has
\begin{equation}\label{supp_ef}{\rm supp\,} e^f(\,\cdot\,, t)
\subset \overline {\Omega^t}\subset \overline
\Omega^T\,,\end{equation} i.e, the trajectory does not leave
$\overline \Omega^T$. Thus, the natural candidate for the role of
the space of states is
$${\cal J}^T:={\rm clos\,}\{y \in \vec L_2(\Omega^T)\,|\,\,{\rm supp\,} y \subset
\Omega^T,\,\,{\rm div\,} y =0\,\,\, {\rm in}\,\,\Omega^T\}\,,$$
where $\rm clos$ is the closure in the vector fields space ${\vec
L_2(\Omega^T)}$ with the product $$(y, v)_{\vec
L_2(\Omega^T)}=\int \limits_{\Omega^T}y \cdot v\,\,\mu$$ and ${\rm
div\,} y=0$ is understood in the sense of distributions. The space
${\cal J}^T$ is called {\it inner}, the waves $e^f(\,\cdot\,, t)$
are its elements.

The inner space contains an extending family of subspaces
$${\cal J}^\xi:={\rm clos}\{y \in {\cal J}^T\,|\,\,{\rm supp\,} y \subset
\Omega^\xi\}\,, \qquad 0 \leq \xi \leq T\,.$$ By the definitions
and (\ref{supp_ef}), one has
\begin{equation}\label{ef_in_Jt}e^f(\,\cdot\,, t) \in {\cal
J}^t\,, \qquad 0 \leq t \leq T\,.\end{equation}
\medskip

\noindent{\bf Control operator}\,\,\,An 'input $\to$ state'
correspondence in the system $\alpha^T$ is realized by a {\it
control operator} $W^T: {\cal F}^T \to {\cal J}^T, \,\,\,{\rm
Dom\,} W^T={\cal F}^T_+$,
$$W^T f\,:=\,e^f(\,\cdot\,, T)\,,$$
which is a closable unbounded operator (see \cite{Eller}). Hence,
it can be extended up to a closed operator, what we assume to be
done\footnote{However, the precise description of ${\rm
Dom\,}\overline W^T$ is not known yet.} and denote the closure by
the same symbol $W^T$.

As each closed operator, the control operator can be represented
in the form of the polar decomposition
\begin{equation}\label{polar}
W^T\,=\,\Phi^T\,|W^T|\,,
\end{equation}
where $|W^T|:=\left[\left(W^T\right)^* W^T\right]^{1 \over 2}$ is
the modulus of $W^T$ and $\Phi^T$ is an isometry from ${\rm Ran\,}
|W^T|$ onto ${\rm Ran\,} W^T$ (see, e.g., \cite{BSol}).

\medskip

\noindent{\bf Response operator}\,\,\, An 'input $\to$ output'
correspondence is described by a {\it response operator} $R^T:
{\cal F}^T \to {\cal F}^T, \,\,{\rm Dom\,} R^T = {\cal M}^T$,
$$\left(R^T f\right)(\gamma, t)\,:=\,\nu (\gamma) \wedge
{\rm curl\,} \int \limits_0^t e^f (\gamma, s)\, ds\,, \qquad
(\gamma, t) \in {{\Gamma \times [0,T]}}\,,$$ where $\nu$ is an
outward normal, the expression in the right hand side being equal
to $-\nu \wedge h^f|_{{\Gamma \times [0,T]}}$. This operator is
unbounded but closable. The latter is a simple consequence of the
following fact: it can be shown that $R^T$ acts continuously from
${\cal F}^T $ to a Sobolev negative space ${\vec
H}^{-p}\left(\Gamma \times [0,T]\right)$ with a big enough $p>0$.

In the mean time, the locality principle enables one to extend the
problem (\ref{Electric1})--(\ref{Electric3}) to the system
\begin{align}
\label{ExtendElectric1}& e_{tt}=-{\rm curl\,}{\rm curl\,} e &&{\rm
in}\,\,D^{2T}\\
\label{ExtendElectric2}& e=0 &&{\rm
in}\,\{(x,t)\in D^{2T}\,|\,t<\tau(x)\} \\
\label{ExtendElectric3}& e_\theta =f &&{\rm on}\,\,\, \Gamma
\times [0,2T]
\end{align}
(see (\ref{Maxextended1})--(\ref{Maxextended3})) and introduce an
{\it extended response operator} $R^{2T}: {\cal F}^{2T} \to {\cal
F}^{2T}, \,\,{\rm Dom\,} R^{2T} = {\cal M}^{2T}$,
$$\left(R^{2T} f\right)(\gamma, t)\,:=\,\nu (\gamma) \wedge
{\rm curl\,} \int \limits_0^t e^f (\gamma, s)\, ds\,, \qquad
(\gamma, t) \in {\Gamma \times [0, 2T]}\,.$$ This operator is also
determined by the part $(\Omega^T,\,g)$ of the manifold and,
hence, can be regarded as an intrinsic object of the system
$\alpha^T$ \footnote{More about the extended (continued) response
operator see in \cite{DSBC}.}.
\medskip

\noindent{\bf Connecting form}\,\,\,A bilinear form $c^T: {\cal
F}^T \times {\cal F}^T \to \mathbb R,\,\,{\rm Dom\,} c^T={\cal
F}^T_+ \times {\cal F}^T_+$,
$$c^T[f,g]\,:=\,\left(e^f(\,\cdot\,, T), e^g(\,\cdot\,, T)\right)_{{\cal J}^T}\,=\,
\left(W^T f, W^T g\right)_{{\cal J}^T}$$ is called {\it
connecting}. By closability of $W^T$, the form $c^T$ is also
closable \cite{BSol}.

The following fact is one of the key points of the BC-method: it
is used in all of its versions \cite{BIP97}, \cite{BIP07}.
\begin{proposition}
The connecting form $c^T$ is determined by the response operator
$R^{2T}$.
\end{proposition}
Moreover, $c^T$ can be expressed through $R^{2T}$ explicitly as
follows. Let $S^T: {\cal F}^T \to {\cal F}^{2T}$ be the operator
that extends the controls, as functions of $t$, from $[0,T]$ to
$[0, 2T]$ by oddness w.r.t. $t=T$. Introduce the class ${\cal
M}^{T,0}:=\left\{f\in {\cal M}^T\,|\,S^T f \in {\cal
M}^{2T}\right\}$; it is dense in ${\cal F}^T_+$ and $S^T {\cal
M}^{T,0} \subset {\rm Dom\,} R^{2T}$ holds. The relation
\begin{equation}\label{cTthroughR2T}
c^T[f,g]\,=\,\left(-\,{1 \over 2}\left(S^T\right)^* R^{2T} S^T
f,\, g\right)_{{\cal F}^T}
\end{equation}
is valid for $f\in {\cal M}^{T,0}, g \in {\cal F}^T_+$ (see
\cite{BGlasAA}).
\begin{corollary}\label{RTTdeterm|WT|}
The response operator $R^{2T}$ determines the operator $|W^T|$.
\end{corollary}
Indeed, $R^{2T}$ determines $c^T$, and one has
\begin{equation}\label{|WT|through_c}
c^T[f,g]\,=\,\left(W^T f,\, W^T g\right)_{{\cal
J}^T}\,=\,\langle{\rm see\,\,}(\ref{polar}) \rangle\,= \left(|W^T|
f,\, |W^T| g\right)_{{\cal F}^T}\,,
\end{equation} whereas the {\it positive} operator $|W^T|$ is determined by
the latter form in the r.h.s..

\subsection{Reachable sets}
Fix an open $\sigma \subset \Gamma$; a linear set
$${\cal U}^\xi_{\rm reach}[\sigma]:=\left\{W^T f\,|\,f \in {\cal M}^T,\,\,
{\rm supp\,} f\subset \sigma \times (T-\xi, T]\right\}$$ is called
{\it reachable} (from $\sigma$, at the time $\xi$). So, the
reachable set consists of electric waves $e^f(\,\cdot\,, T)$
produced by all smooth delayed controls, which act from the part
of boundary $\sigma$, the action time being equal to $\xi$. Its
closure (in ${\cal J}^T$)
$${\cal U}^\xi[\sigma]\,:=\,{\rm clos\,} {\cal U}^\xi_{\rm reach}[\sigma]$$ is
said to be a {\it reachable subspace}. Also, we denote shortly
${\cal U}^\xi[\Gamma]=:{\cal U}^\xi$.

For a measurable subset $A \subset \Omega^T$, we define a subspace
$${\cal J} \langle A \rangle\,:=\,{\rm clos\,} \{y \in {\cal J}^T\,|\, {\rm supp\,} y \subset
A\}\,.$$ By the finiteness of the wave speed propagation, each
wave belonging to ${\cal U}^\xi_{\rm reach}[\sigma]$ is supported
in the metric neighborhood ${\Omega}^\xi[\sigma]$. Hence, the
embedding
\begin{equation}\label{U[sigma]subsetJ[sigma]}
{\cal U}^\xi[\sigma] \subset {\cal
J}\langle\Omega^\xi[\sigma]\rangle
\end{equation}
holds. A structure of the reachable sets and subspaces, as well as
the character of the embedding (\ref{U[sigma]subsetJ[sigma]}) is
the subject of the boundary control theory. For our goals, the
following fact is of crucial value.
\begin{proposition}
Any field $h \in {\cal J}\langle\Omega^\xi[\sigma]\rangle \ominus
{\cal U}^\xi[\sigma]$ is smooth and satisfies
\begin{equation}\label{curlh=0}
{\rm curl\,} h\,=\,0
\end{equation}
in ${\rm int\,}\Omega^\xi[\sigma]$. \end{proposition} {\bf
Proof}\,\,\, see in \cite{BGlasAA}, \cite{BIP07} and the remark in
the third item of sec 0.3; more about properties of the reachable
sets see in \cite{BGlasCOCV}. Note that the derivation of
(\ref{curlh=0}) relays upon the fundamental Holmgren-John-Tataru
theorem on uniqueness of continuation of the solutions to the
Maxwell system across a non-characteristic surface \cite{EINT}.
\begin{corollary}\label{J<omega>subsUxisigma}
If a subdomain $\omega \subset {\rm int\,}\Omega^\xi[\sigma]$ is
homeomorphic to an open ball in ${\mathbb R}^3$, then the relation
\begin{equation}\label{embedding}
{\cal J}\langle\omega\rangle\, \subset \,{\cal U}^\xi[\sigma]
\end{equation}
holds.
\end{corollary}
Indeed, by (\ref{curlh=0}) any $h \in {\cal
J}\langle\Omega^\xi[\sigma]\rangle \ominus {\cal U}^\xi[\sigma]$
can be represented locally as $h|_{\omega}=\nabla p$. Hence, for
$y \in {\cal J}\langle\omega\rangle \subset {\cal
J}\langle\Omega^\xi[\sigma]\rangle$ one has
\begin{equation*}
(h, y)_{{\cal J}^T}=\int \limits_\omega \nabla p \cdot y\,\,\mu
=\langle {\rm see\,} (\ref{Green})\rangle\,=\,0\,,
\end{equation*}
i.e., $y \in  {\cal U}^\xi[\sigma]$ and we arrive at
(\ref{embedding}). This result shows that the reachable subspace
${\cal U}^\xi[\sigma]$ is always rich enough. Moreover, in a
relevant sense, the generic case is that the defect subspace
${\cal J}\langle\Omega^\xi[\sigma]\rangle \ominus {\cal
U}^\xi[\sigma]$ is finite dimensional, its dimension being
determined by topology of $\Omega^\xi[\sigma]$ (see, e.g.,
\cite{BGlasCOCV}).

\section{Inverse problem}
\subsection{Wave caps}
Here the main technical tool for solving the inverse problem is
introduced.

In parallel with the definition (\ref{cap}), introduce a subspace
\begin{equation}\label{wavecap} w^{s,\varepsilon}_\gamma\,:={\cal
U}^s[\sigma_\varepsilon(\gamma)] \bigcap \left\{{\cal U}^s\ominus
{\cal U}^{s-\varepsilon}\right\}\end{equation} and call it a {\it
wave cap}. The correspondence between the wave and 'space' caps is
as follows. Assume that $\gamma, \varepsilon$, and $s$ are chosen
so that the cap $\omega^{s,\varepsilon}_\gamma$ is nonempty.
\begin{itemize}
\item {\it Each $y \in w^{s,\varepsilon}_\gamma$ is supported in
$\overline \omega^{s,\varepsilon}_\gamma$.} Indeed, by
(\ref{curlh=0}) with $\sigma=\Gamma$, the orthogonality $y \bot\,
{\cal U}^{s-\varepsilon}$ implies that the field $y$ is
harmonic\footnote{i.e., ${\rm curl\,} y=0$ and ${\rm div\,} y=0$;
see, e.g., \cite{Schw}} in $\Omega^{s-\varepsilon}$. In the mean
time, $y$ is supported in the neighborhood
$\overline\Omega^s[\sigma_\varepsilon(\gamma)]$ and, whence,
vanishes out of this neighborhood. By the uniqueness of the
harmonic continuation, $y$ has to vanish everywhere in
$\Omega^{s-\varepsilon}$ and, thus, to be supported in the space
cap. \item Fix an $x$ and a (small) ball $\Omega^\delta[x]\subset
{\rm int\,}\omega^{s,\varepsilon}_\gamma$. As is easy to see from
(\ref{embedding}), the subspace ${\cal
J}\langle\Omega^\delta[x]\rangle$ is embedded in the r.h.s. of
(\ref{wavecap}). Hence, {\it the embedding ${\cal
J}\langle\Omega^\delta[x]\rangle \subset w^{s,\varepsilon}_\gamma$
holds} and shows that the wave cap is a rich enough nonzero
subspace.
\end{itemize}
As one can easily conclude from the above-mentioned facts, the
space and wave caps behave in one and the same way: the
equivalence
\begin{equation}\label{wcap=scap}
\left\{\omega^{s,\varepsilon}_\gamma\neq\emptyset\right\}\Longleftrightarrow
\left\{w^{s,\varepsilon}_\gamma\neq\{0\}\right\}
\end{equation}
holds.

Return to Corollaries 1--3. The 'subdomain $\leftrightarrow$
subspace' arguments quite analogous to the ones that have led to
(\ref{wcap=scap}) imply  the equivalences
\begin{align}
\label{|A|} & \left\{\omega^{s,\varepsilon}_{\gamma'} \bigcap
\,\Omega^{r+\varepsilon}[\sigma_\varepsilon(\gamma)]\,\not=\,\emptyset\right\}
\Longleftrightarrow \left\{w^{s,\varepsilon}_{\gamma'} \bigcap
\,{\cal
U}^{r+\varepsilon}[\sigma_\varepsilon(\gamma)]\,\not=\,\{0\}\right\}\\
\label{|B|} & \left\{\omega^{s,\varepsilon}_\gamma \bigcap\,
\Omega^{s+\varepsilon}\left[
\sigma_\varepsilon({\gamma'})\right]\not=\emptyset\right\}\Longleftrightarrow
\left\{w^{s,\varepsilon}_\gamma \bigcap\, {\cal
U}^{s+\varepsilon}\left[
\sigma_\varepsilon({\gamma'})\right]\not=\{0\}\right\}\,.
\end{align}

\subsection{Model caps}
Return to the polar decomposition (\ref{polar}).

Fix an open $\sigma \subset \Gamma$. In the outer space ${\cal
F}^T$, introduce a linear set
\begin{align}\notag & |{\cal U}|^\xi_{\rm reach}[\sigma]\,:=\,
\left\{|W|^T f\,|\,\,f \in {\cal M}^T,\,\, {\rm supp\,} f\subset
\sigma \times (T-\xi, T]\right\}\\ \label{|Ureach|} &
=\,(\Phi^T)^*\,{\cal U}^\xi_{\rm reach}[\sigma]
\end{align} and a subspace
\begin{align}\label{|U|} |{\cal U}|^\xi[\sigma]\,:={\rm clos\,} |{\cal U}|^\xi_{\rm reach}[\sigma]\,=\,(\Phi^T)^*\, {\cal U}^\xi[\sigma]
\end{align}
(the closure in ${\cal F}^T$) \,that we call the {\it model}
reachable set and subspace respectively. Subsequently, define a
{\it model wave cap}
\begin{equation}\label{|wavecap|}|w|^{s,\varepsilon}_\gamma\,:=\,
|{\cal U}|^s[\sigma_\varepsilon(\gamma)] \bigcap \left\{|{\cal
U}|^s\ominus |{\cal U}|^{s-\varepsilon}\right\}\,=\,(\Phi^T)^*\,
w^{s,\varepsilon}_\gamma.\end{equation} Since the map $(\Phi^T)^*$
is an isometry, the subspaces in the right hand sides of the
equivalences (\ref{|A|}) and (\ref{|B|}) can be replaced by their
model copies, i.e., the images through this map. Indeed, for
instance, by virtue of the evident equality
$$|w|^{s,\varepsilon}_{\gamma'} \bigcap
\,|{\cal
U}|^{r+\varepsilon}[\sigma_\varepsilon(\gamma)]\,=\,(\Phi^T)^*\,\left(w^{s,\varepsilon}_{\gamma'}
\bigcap \,{\cal
U}^{r+\varepsilon}[\sigma_\varepsilon(\gamma)]\right)$$ the
conditions $$w^{s,\varepsilon}_{\gamma'} \bigcap \,{\cal
U}^{r+\varepsilon}[\sigma_\varepsilon(\gamma)]\,\not=\,\{0\}
\qquad {\rm and} \qquad |w|^{s,\varepsilon}_{\gamma'} \bigcap
\,|{\cal
U}|^{r+\varepsilon}[\sigma_\varepsilon(\gamma)]\,\not=\,\{0\}$$
are equivalent.

Return once again to Corollaries 1--3. The aforesaid enables one
to reformulate them as follows.
\begin{corollary}\label{corollary|1|}
Let $T>0$ be fixed. A point $(\gamma,s)\in \Gamma \times [0,T)$
belongs to the set $\Theta^T \cup \theta^T$ iff for any
$\varepsilon>0$ the relation
\begin{equation}\label{|recoverpattern|}
|w|^{s,\varepsilon}_\gamma \not=\{0\}
\end{equation}
holds; in this case, the inequality $s\leq \tau_*(\gamma)$ is
valid. Otherwise, if the family of caps terminates, one has
$(\gamma,s)\notin \Theta^T \cup \theta^T$ and, hence, $s>
\tau_*(\gamma)$ is valid.
\end{corollary}
\begin{corollary}\label{corollary|2|}
Let $\gamma \in \Gamma$ and $(\gamma',s) \in \Theta^T$, so that
$x'=x(\gamma',s) \in \Omega^T \backslash c$. For a fixed $r<T$,
the inclusion $x' \in \overline \Omega^r[\gamma]$ in $\Omega^T$
(or, equivalently, the inclusion $(\gamma',s) \in i\left(\overline
\Omega^r[\gamma]\backslash c\right)$ on $\Theta^T$) holds iff the
relation
\begin{equation}\label{|recoverspheres|}
|w|^{s,\varepsilon}_{\gamma'} \bigcap \,|{\cal
U}|^{r+\varepsilon}[\sigma_\varepsilon(\gamma)]\,\not=\,\{0\}
\end{equation}
holds for any $\varepsilon>0$.
\end{corollary}
\begin{corollary}\label{corollary|3|}
 Let the points $(\gamma,s)$ and
$(\gamma',s)$ belong to the coast $\theta^T$. The relation
$(\gamma,s) \overset{E}= (\gamma',s)$ is valid iff for any
$\varepsilon>0$ the relation
\begin{equation}\label{|gluingpattern|}|w|^{s,\varepsilon}_\gamma \bigcap\, |{\cal U}|^{s+\varepsilon}\left[
\sigma_\varepsilon({\gamma'})\right]\not=\{0\}
\end{equation}
holds (or, equivalently, $|w|^{s,\varepsilon}_{\gamma'} \cap
|{\cal
U}|^{s+\varepsilon}\left[\sigma_\varepsilon({\gamma})\right]\not=\{0\}$).
\end{corollary}
The key fact is that the subspaces in the l.h.s. of
(\ref{|recoverpattern|})--(\ref{|gluingpattern|}) are determined
by the operator $|W^T|$ and, whence, by the inverse data (operator
$R^{2T}$).

\subsection{Reconstruction}
Now, to solve the inverse problem it suffices just to summarize
our previous considerations, which we present in the form of the
following procedure. Recall the starting point: {\it we are given
with the extended response operator $R^{2T}$}.
\smallskip

\noindent{\bf Step 1}\,\,\,Recover the connecting form $c^T$ from
(\ref{cTthroughR2T}) and determine the operator $|W^T|$ (see
Corollary \ref{RTTdeterm|WT|}, (\ref{|WT|through_c})).
\smallskip

\noindent{\bf Step 2}\,\,\,Determine the sets $|{\cal U}|^\xi_{\rm
reach}[\sigma]$, subspaces $|{\cal U}|^\xi[\sigma]$, and the model
caps $|w|^{s,\varepsilon}_{\gamma}$ by the definitions
(\ref{|Ureach|}), (\ref{|U|}), and (\ref{|wavecap|}) respectively
for those $\sigma, \varepsilon, \gamma, \xi$, which are used in
Corollaries \ref{corollary|1|}--\ref{corollary|3|}.
\smallskip

\noindent{\bf Step 3}\,\,\,Checking the relation
(\ref{|recoverpattern|}), recover the pattern $\Theta^T$ and its
coast $\theta^T$.

\noindent{\bf Step 4}\,\,\, By checking (\ref{|recoverspheres|}),
determine the family of ball images ${\cal B}^T$ and recover the
metric tensor $g_{\rm sgc}$ on the pattern (see Lemma
\ref{Lemma1}).
\smallskip

\noindent{\bf Step 5}\,\,\,By the use of (\ref{|gluingpattern|}),
recover the equivalence $E$. The collection $\Theta^T, \theta^T,
E$ determines the manifold $({\widetilde \Omega}^T, \,\widetilde
g)$, which can be constructed by means of the procedure described
in sec 1.5.
\smallskip

As a result, the operator $R^{2T}$ determines the manifold
$({\widetilde \Omega}^T, \,\widetilde g)$, which is isometric to
the manifold $(\Omega^T,\,g)$ by construction. In other words, it
determines $(\Omega^T,\,g)$ up to isometry that proves Theorem 1.
Moreover, identifying the points by
$$\Omega^T \supset \Gamma \ni \gamma\,\equiv\,\pi\left((\gamma,
0)\right) \in
\partial \widetilde\Omega^T \subset \widetilde\Omega^T$$ we can easily check that the response operator
$\widetilde R^{2T}$ of the constructed manifold is identical to
$R^{2T}$. The latter motivates to refer to $({\widetilde
\Omega}^T, \,\widetilde g)$ as a {\it canonical representative} of
the class of isometric manifolds possessing the given inverse data
$R^{2T}$.
\smallskip

In conclusion, note the following. It is not inconvincible that
the above described procedure $R^{2T} \Rightarrow ({\widetilde
\Omega}^T, \,\widetilde g)$ is available for numerical
realization. The principal problem (and difficulty) is to get good
enough simulation of the wave caps $w^{s,\varepsilon}_{\gamma}$
that is to provide a rich enough set of waves $e^f(\,\cdot\,, T)$
concentrated near the point $x(\gamma, s)$. The sharper is the
concentration, the better is the reconstruction of the pattern,
distant functions $r_a$, etc. Such a principle is traced in all
versions of the BC-method.


\begin{thebibliography}{15}
\bibitem{BIP97}
M.I.~Belishev.
\newblock {Boundary control in reconstruction of manifolds and
metrics (the BC method).}
\newblock {\em Inverse Problems}, 13(5): R1--R45, 1997.

\bibitem{DAN}
M.I.Belishev, V.M.Isakov, L.N.Pestov, and V.A.Sharafutdinov.
\newblock{On the Reconstruction of Metrics via External Electromagnetic Measurements.}
\newblock{\em \newblock Doklady Mathematics}, 61 (2000), No 3, 353--356.

\bibitem{BGlasAA}
M.I.Belishev, A.K.Glasman.
\newblock{Dynamical inverse problem for the Maxwell system: recovering the
velocity in a regular zone (the BC--method).}
\newblock{\em St.-Petersburg Math. Journal}, 12 (2001), No 2: 279--316.

\bibitem{BGlasCOCV}
M.I.Belishev and A.K.Glasman.
\newblock{Boundary control of the Maxwell dynamical system: lack of
controllability by topological reasons.}
\newblock{\em \newblock ESAIM COCV}, 5 (2000),
207--217.

\bibitem{DSBC}
M.I.Belishev.
\newblock{Dynamical systems with boundary control: models and
characterization of inverse data.}
\newblock{\em Inverse Problems}, 17 (2001), 659--682.

\bibitem{BIP07}
M.I.Belishev.
\newblock {Recent progress in the boundary control method.}
\newblock {\em Inverse Problems}, 23 (2007), no 5, R1--R67.

\bibitem{BSol}
M.S.~Birman, M.Z.~Solomyak.
\newblock {Spectral Theory of Self-Adjoint Operators in Hilbert Space.}
\newblock {\em D.Reidel Publishing Comp.}, 1987.

\bibitem{EINT}
M.Eller, V.Isakov, G.Nakamura, D.Tataru.
\newblock {Uniqueness and stability in the Cauchy problem for
Maxwell's and elasticity systems.}
\newblock {\em Nonlinear PDE and Applications,
Eds. D.Cioranescu, J-L. Lions, College de France Seminar}, 14,
329--349,
\newblock{"Studies in Mathematics and its applications"},
v.31, North-Holland, Elsevier Science, 2002.

\bibitem{Eller}
M.Eller.
\newblock {Symmetric hyperbolic systems with boundary
conditions that do not satisfy the Kreiss–-Sakamoto condition.}
\newblock {\em Applicationes Mathematicae}, 35 (2008), 323-333.

\bibitem{GKM}
D.Gromol, W.Klingenberg, W.Meyer.
\newblock{Riemannische Geometrie im Grossen},
\newblock{\em Berlin, Springer}, 1968.

\bibitem{Isbook}
V.Isakov.
\newblock{Inverse problems for partial differential equations.}
\newblock{\em Appl. Math. Studies, Springer}, v. 127, 2006.

\bibitem{Kling}
W.Klingenberg. \newblock{Riemannian geometry.}
\newblock {\em Series YB de Gruyer: Studies in Mathematics},
vol 1, 1982.

\bibitem{Rom1}
V.G.Romanov.
\newblock {On solvability of the inverse problems for the hyperbolic equations
in a class of functions analytical with respect to a part of
variables.}
\newblock {\em Doklady Akad. Nauk SSSR}, 304 (4): 51--67, 1989
(in Russian).

\bibitem{Rom2}
V.G.Romanov.
\newblock {Stability in inverse problems.}
\newblock {\em Moskva, Nauchnyi Mir}, 2005 (in Russian).

\bibitem{RomPukh}
V.G.Romanov, T.N.Pukhnacheva.
\newblock {Solution stability theorem in coefficients determination problem for
the Maxwell equations system.}
\newblock {\em Ill-Posed Problems; Ed. M.M.Lavrent'ev},
\newblock {\em Novosibirsk, SO AN SSSR}, 1988, 64--86 (in Russian).

\bibitem{Schw}
G.Schwarz.
\newblock {Hodge decomposition - a method for solving boundary value
problems.}
\newblock {\em Lecture notes in Math.}, 1607.
\newblock{Springer--Verlag, Berlin}, 1995.
\end{thebibliography}
\end{document}